\documentclass[a4paper,superscriptaddress,floatfix]{revtex4}

\usepackage{graphicx,multirow}
\usepackage{bm}
\usepackage{amsmath}
\usepackage{amssymb}
\usepackage{amscd}
\usepackage{latexsym}
\usepackage{slashed}
\usepackage{color}
\usepackage{graphicx}
\usepackage{ulem}
\usepackage{color}
\usepackage[caption=false]{subfig}
\usepackage{mathtools}
\usepackage{xspace}

\newcommand{\tanb}         {\ensuremath{\tan\beta}\xspace}

\newcommand{\GeV}{\ensuremath{\mathrm{GeV}}\xspace}
\newcommand{\TeV}{\ensuremath{\mathrm{TeV}}\xspace}

\begin{document}

\title{Below-threshold CP-odd Higgs boson search via $A\to Z^* h$ at the LHC}

\author{E. Accomando}%
 \email{e.accomando@soton.ac.uk}
\affiliation{School of Physics and Astronomy, University of Southampton, Highfield, Southampton SO17 1BJ, UK}%
\author{M. Chapman}%
\email{matthew.chapman@soton.ac.uk}
\affiliation{School of Physics and Astronomy, University of Southampton, Highfield, Southampton SO17 1BJ, UK}%
\affiliation{H.H. Wills Physics Laboratory, University of Bristol, Bristol, BS8 1TL, UK}%
\author{A. Maury}%
\email{arnaud.maury@universite-paris-saclay.fr}
\affiliation{Magist{\`{e}}re de Physique Fondamentale, Universit{\'{e}} Paris-Saclay, Batiment 625, 91 405 Orsay, France}%
\author{S. Moretti}%
 \email{s.moretti@soton.ac.uk}
\affiliation{School of Physics and Astronomy, University of Southampton, Highfield, Southampton SO17 1BJ, UK}%

\begin{abstract}
\noindent
We study the process $q\bar q, gg \to A\to Z^* h$ in a 2-Higgs Doublet Model Type-II where the mass of the CP-odd Higgs state $A$ is lower than the rest mass of the $Z$ and $h$ particles (the latter being the Standard Model-like Higgs state discovered at the Large Hadron Collider in 2012), i.e., $m_A<m_Z+m_h\approx 215$ GeV. This is a mass range which is not being currently tested by ATLAS and CMS, yet we  show that there can be sensitivity to it already during Runs 2 \& 3, assuming leptonic decays of the gauge boson and bottom-antibottom quark ones for the Higgs boson.
\end{abstract}

\maketitle

\section{Introduction}
\noindent
The Higgs boson predicted by the Standard Model (SM), hereafter denoted as $h$, appears to have been discovered at the Large Hadron Collider (LHC) by both the ATLAS and CMS collaborations, back in July 2012 \cite{Aad:2012tfa,Chatrchyan:2012xdj}. Indeed, all its properties (Yukawa and gauge couplings, spin-0, CP-even state) seem consistent with those of the SM state. Yet, the so-called SM (or alignment) limit exists in a variety of Beyond the SM (BSM) scenarios. One amongst the latter that deserves particular attention is probably the 2-Higgs Doublet Model (2HDM) \cite{HiggsHunters,Branco2012}. This is because it is the simplest generalisation of the SM that relies exclusively on the only  Higgs  multiplet structure so far revealed by Nature, i.e., a doublet one. However, it adds to the SM some notable features, specifically, that its particle content includes all possible other states that Nature already incorporates in its gauge  sector, as far as mass, electro-magnetic charges and CP quantum numbers are concerned. These are the $A$  (massive, neutral and CP-odd, like the $Z$)  and $H^\pm$ (massive, charged and with mixed CP, like the $W^\pm$) Higgs states. In fact, somewhat of a redundancy, there is also a CP-even companion to the $h$ state in the 2HDM, the $H$ one, which we assume to be heavier in comparison (i.e., $m_H>m_h$). Intriguingly then, this is an appealing scenario to search for, further considering the fact that the 2HDM can naturally comply with the stringent limits from Electro-Weak Precision Observables (EWPOs), as it suffices to set the $H^\pm$ state somewhat degenerate in mass with either the $A$ or $H$ ones. Furthermore,  it embeds the aforementioned alignment limit, which is obtained when $\cos(\beta-\alpha)=1$  (where $\alpha$ is the mixing angle between the $h$ and $H$ states and $\tan\beta$ is the ratio of the Vacuum Expectation Values (VEVs) of the two doublets). Finally, it can avoid large Flavour Changing Neutral Currents (FCNCs) by simply invoking a $\mathbb{Z}_2$ symmetry between the two Higgs doublet fields, which can in fact be softly broken through a mass parameter (denoted by $m_{12}$) mixing the two Higgs doublets 
without falling foul of current experimental limits. {Note that $m_h$, $m_H$, $m_A$, $m_{H^\pm}$, $\alpha$, $\tan\beta$ and $m_{12}$ can be taken as the seven independent parameters of the 2HDM, though other choices are possible. In this letter, we explicitly consider a CP-conserving scenario within the 2HDM.}

There are innumerable searches that have been carried out at the LHC looking for new Higgs bosons. Here, we concentrate on those for the $A$ state. Amongst these, we select the one which tried to extract an $A\to Zh$ signal. This particularly attractive mode enables the reconstruction of both the $Z$ and $h$ masses, which are well measured at 91 and 125 GeV, respectively, e.g., by using the  $Z\to l^+l^-$ ($l=e,\mu$) and $h\to b\bar b$ decays. This was the signature exploited by the ATLAS analysis of Ref.~\cite{TheATLAScollaboration:2016loc} (see also 
\cite{Pandini:2016utq}). However, the latter (like all previous studies {in the $Zh$ channel possibly mediated by the CP-odd Higgs}) concentrated on an $A$ mass range starting from $m_Z+m_h \approx$ 215 GeV, i.e., assuming decays of the CP-odd state into $Z$ and $h$ particles  both being on-shell. While this assumption is fully justified in the case of the Higgs boson, which has a width of order 10 MeV (i.e., $0.03\%$ of its mass), it is less so for the gauge boson, for which the width-to-mass ratio is of order $3\%$. {Off-shell effects involving the $Z$ boson are therefore not negligible, hence searching for the CP-odd Higgs boson decaying into $Z^*h$ is of phenomenological interest.}
Therefore, in this paper, we ask ourselves the question of which region of parameter space can be accessed at the LHC, in the context of the 2HDM, if one looks instead for $A\to Z^* h$ decays, wherein the $Z$ boson is off-shell. In the next section, we discuss the 2HDM. In Sect. III we present our results. We then conclude.


\section{The 2HDM}

\noindent
In this section we give a brief introduction to the 2HDM, with a focus on the aspects relevant to our analysis. Extensive reviews can be found in
Refs.~\cite{HiggsHunters,Branco2012}. As intimated, in the 2HDM, one extends the SM Higgs sector by including two complex doublets, which eventually give rise to a spectrum containing five physical Higgs states, $h$, $H$, $A$ and $H^{\pm}$. 

\noindent
{The most general renormalisable (i.e.,  quartic) scalar potential with two Higgs doublets
can be written as}
\begin{equation}
\begin{aligned}
	\mathcal{V}_{gen}
	&=
	m_{11}^{2} \Phi_{1}^{\dag} \Phi_{1} 
	+ m_{22}^{2} \Phi_{2}^{\dag} \Phi_{2}
	- \left[ m_{12}^{2} \Phi_{1}^{\dag} \Phi_{2} + \rm{h.c.} \right] + \\
	&+ \frac{1}{2} \lambda_{1} \left( \Phi_{1}^{\dag} \Phi_{1} \right)^{2}
	+ \frac{1}{2} \lambda_{2} \left( \Phi_{2}^{\dag} \Phi_{2} \right)^{2}
	+ \lambda_{3} \left( \Phi_{1}^{\dag} \Phi_{1} \right) \left( \Phi_{2}^{\dag} \Phi_{2} \right)
	+ \lambda_{4} \left( \Phi_{1}^{\dag} \Phi_{2} \right) \left( \Phi_{2}^{\dag}
	\Phi_{1} \right) + \\
	&+ \left\{ \frac{1}{2} \lambda_{5} \left( \Phi_{1}^{\dag} \Phi_{2} \right)^{2} + \left[ \lambda_{6} \left( \Phi_{1}^{\dag} \Phi_{1} \right) + \lambda_{7} \left( \Phi_{2}^{\dag} \Phi_{2} \right) \right] \Phi_{1}^{\dag} \Phi_{2} + \rm{h.c.} \right\},
\end{aligned}
\end{equation}
\noindent
{where $m_{11}^{2}$, $m_{22}^{2}$, $m_{12}^{2}$ are mass squared parameters and 
$\lambda_{i}$ ($i = 1, ..., 7$) are dimensionless quantities describing the
coupling of the order-4 interactions. Six parameters are real ($m_{11}^{2}$,
$m_{22}^{2}$, $\lambda_{i}$ with $i = 1, ..., 4$) and four are a priori complex
($m_{12}^{2}$ and $\lambda_{i}$ with $i = 5, ..., 7$). Therefore, in general,
the model has 14 free parameters. Under appropriate constraints, this number
can however be reduced.}

\noindent
{The potential is explicitly CP-conserving if and only if there exists a basis choice for the scalar fields in which $m_{12}^{2}$, $\lambda_{5}$, $\lambda_{6}$ and $\lambda_{7}$ are real. In this letter, we assume that both the scalar potential and the vacuum are CP-conserving. Consequently, the number of free parameters goes down to 10. After ElectroWeak  Symmetry Breaking (EWSB), each scalar doublet acquires a VEV that can be parametrised as follows:}
\begin{equation}
  \langle \Phi_{1} \rangle =
	\frac{v}{\sqrt{2}}
  \begin{pmatrix}  
	0 \\
	\cos\beta
  \end{pmatrix}
\quad
\quad
  \langle \Phi_{2} \rangle =
	\frac{v}{\sqrt{2}}
  \begin{pmatrix}  
	0 \\
	\sin\beta
  \end{pmatrix},
\end{equation}
\noindent
{where the angle $\beta$ determines the ratio of the two doublet VEVs, $v_1$ and $v_2$, through the definition of $\tan\beta = v_{2} / v_{1}$. The angle $\beta$ is an additional parameter that adds to the free parameters defining the scalar potential.}

{
\begin{table}[t!]
\begin{center}
\begin{tabular}{|c|c|c|c|c|c|c|c|c|c|}
   \hline
	\multirow{2}{*}{Model} & \multicolumn{3}{c|}{$h$} & \multicolumn{3}{c|}{$H$} & \multicolumn{3}{c|}{$A$} \tabularnewline
	\cline{2-10}
	 &
	$u$ & $d$ & $l$ &
	$u$ & $d$ & $l$ &
	$u$ & $d$ & $l$
	\tabularnewline
   \hline
	{Type-I}  &
	$\phantom{-}\frac{\cos\alpha}{\sin\beta}$ & $\phantom{-}\frac{\cos\alpha}{\sin\beta}$ & $\phantom{-}\frac{\cos\alpha}{\sin\beta}$ & 
	$\phantom{-}\frac{\sin\alpha}{\sin\beta}$ & $\phantom{-}\frac{\sin\alpha}{\sin\beta}$ & $\phantom{-}\frac{\sin\alpha}{\sin\beta}$ & 
	$\phantom{-}\cot\beta                   $ & $         - \cot\beta                   $ & $         - \cot\beta                $         
	\tabularnewline
   \hline
	{Type-II}  &
	$\phantom{-}\frac{\cos\alpha}{\sin\beta}$ & $          -\frac{\sin\alpha}{\cos\beta} $ & $          -\frac{\sin\alpha}{\cos\beta} $ & 
	$\phantom{-}\frac{\sin\alpha}{\sin\beta}$ & $\phantom{-}\frac{\cos\alpha}{\cos\beta} $ & $\phantom{-}\frac{\cos\alpha}{\cos\beta} $ & 
	$            \cot\beta                  $ & $           \tan\beta                    $ & $           \tan\beta                    $   
	\tabularnewline
   \hline
	{Type-III}  &
	$\phantom{-}\frac{\cos\alpha}{\sin\beta}$ & $\phantom{-}\frac{\cos\alpha}{\sin\beta}$  & $           -\frac{\sin\alpha}{\cos\beta}$ & 
	$\phantom{-}\frac{\sin\alpha}{\sin\beta}$ & $\phantom{-}\frac{\sin\alpha}{\sin\beta} $ & $\phantom{-}\frac{\cos\alpha}{\cos\beta} $ & 
	$          \cot\beta                   $ & $\phantom{-}-\cot\beta                    $ & $           \tan\beta                   $   
	\tabularnewline
   \hline
	{Type-IV}  &
	$\phantom{-}\frac{\cos\alpha}{\sin\beta}$ & $           -\frac{\sin\alpha}{\cos\beta}$ & $\phantom{-}\frac{\cos\alpha}{\sin\beta} $ & 
	$\phantom{-}\frac{\sin\alpha}{\sin\beta}$ & $\phantom{-}\frac{\cos\alpha}{\cos\beta} $ & $\phantom{-}\frac{\sin\alpha}{\sin\beta} $ & 
	$           \cot\beta                  $ & $           \tan\beta                   $ & $\phantom{-}-\cot\beta                    $   
	\tabularnewline
   \hline
\end{tabular}
\caption{Couplings of the neutral Higgs bosons to fermions, normalised to the corresponding SM value ($m_{f}/v$, $f=u,d,l$) in the 2HDM Type-I, -II, -III and -IV. The $A\bar f f$ couplings are meant for $iA\bar f\gamma_5f$.}
\label{tab:interaction}
\end{center}
\end{table}
}

\noindent
{In general, the Yukawa matrices corresponding to the two Higgs doublets cannot be simultaneously
diagonalised, which can pose a problem, as the off-diagonal elements lead to tree-level Higgs mediated FCNCs on which severe experimental bounds exist. The Glashow-Weinberg-Paschos (GWP) theorem~\cite{GWP1,GWP2} states that this type of FCNCs is absent  if at most one Higgs multiplet is responsible for providing mass to fermions of a given electric charge. This GWP condition can be enforced by a discrete 
$\mathbb{Z}_2$ symmetry ($\Phi_1 \rightarrow +\Phi_1$ and $\Phi_2\rightarrow -\Phi_2$) on the doublets, in which case the absence of FCNCs is natural. The soft $\mathbb{Z}_2$ breaking condition relies on the existence of a basis where $\lambda_6 = \lambda_7$ = 0. Therefore, one looses two additional degrees of freedom reducing the number of free parameters down to 9. Finally, $m_{11}^{2}$ and $m_{22}^{2}$ can be expressed as a function of the other parameters, owing to the fact that the scalar potential is in a local minimum when computed in the VEVs. So, globally, with restrictions to CP-conservation and soft $\mathbb{Z}_2$  symmetry breaking, there remain seven free parameters in the 2HDM.}

\noindent
{There exist several alternative bases in which the 2HDM can be described. The scalar potential given above in terms of the $m^2_{ij}$'s and $\lambda_i$'s defines the so-called {\it general parametrisation}.} However, it is customary to parameterise this scenario by using the \textit{hybrid basis} of Ref.\cite{Haber2015}, where the parameters provide a convenient choice to give a direct control on both the CP-even and CP-odd Higgs boson masses, the $hVV$ couplings ($V$ = $W^\pm , Z$), the $Af\bar{f}$ vertices (where $f$ is a fermion) and the Higgs quartic couplings. The parameters in this basis are:
\begin{equation}
	m_{h}, \quad
	m_{H}, \quad
        \cos(\beta - \alpha), \quad
	\tan\beta , \quad
	Z_{4}, \quad
	Z_{5}, \quad
	Z_{7},
\end{equation}
\noindent
with the CP-even Higgs boson masses satisfying $m_H\ge m_h$ and the angles being $0\leq\beta\leq\pi /2$ and $0\le\sin (\beta -\alpha )\leq1$. The $\cos(\beta - \alpha )$ parameter determines the couplings of the CP-even Higgs bosons with the SM gauge bosons, $g_{hVV}$ and $g_{HVV}$ ($V = W^\pm, Z$), and the CP-odd Higgs boson $g_{AhZ}$. The latter is of particular interest in this analysis. The $Z_{4, 5, 7}$ parameters are instead the Higgs self-couplings. They are given by
\begin{equation}
Z_{4} = {1\over 4}\sin^2(2\beta )[\lambda_1+\lambda_2-2(\lambda_3 + \lambda_4 + \lambda_5)]+\lambda_{4}
\end{equation}
\begin{equation}
Z_5 = {1\over 4}\sin^2(2\beta )[\lambda_1+\lambda_2-2(\lambda_3 + \lambda_4 + \lambda_5)] + \lambda_5
\end{equation}
\begin{equation}
Z_7 = -{1\over 2}\sin (2\beta )[\lambda_1\sin^2(\beta ) -\lambda_2\cos^2(\beta ) +(\lambda_3 + \lambda_4 + \lambda_5)\cos(2\beta )]
\end{equation}
\noindent
The remaining  Higgs boson masses can be expressed in terms of the quartic scalar couplings $Z_4$ and $Z_5$:
\begin{equation}
	m_A^2 = m_H^2\sin^2(\beta -\alpha) + m_h^2\cos^2(\beta -\alpha) - Z_5v^2,
\end{equation}
\begin{equation}
	m_{H^\pm}^2 = m_A^2 -{1\over 2}(Z_4 - Z_5)v^2.
\end{equation}

In the hybrid basis, swapping the self-couplings $Z_4$ and $Z_5$ with the scalar masses given above, the seven free parameters can be recast into four physical masses and three parameters that are related to the couplings of the scalars to gauge bosons, fermions and themselves, respectively:
\begin{equation}
	m_h, ~ m_H, ~ m_A, ~ m_{H^\pm}, ~ \cos (\beta - \alpha ), ~ \tan (\beta ), ~ Z_7.
\label{parameters}
\end{equation}

\noindent
In Eq. (\ref{parameters}), $Z_7$ enters only the Higgs triple and quartic  interactions. It plays an important role also in setting the minimum CP-odd Higgs mass allowed by the theoretical constraints of perturbativity and stability. Beside the Higgs fields, also fermions are required to have a definite charge under the discrete $\mathbb{Z}_2$ symmetry. The different assignments of the $\mathbb{Z}_2$ charge in the fermion sector give rise to  four different types of 2HDM. The couplings of the neutral Higgses to fermions, normalised to the corresponding SM value ($m_{q}/v$) (henceforth, denoted by $\kappa_{hqq}$ for the case of the SM-like Higgs state coupling to a quark $q$,  where $q=d,u$), can be found in Tab.~\ref{tab:interaction}. 

\noindent
In the remainder of this paper, we will concentrate on the 2HDM Type-II. Herein, there are two limiting scenarios, giving rise to two distinct regions in the $(\cos (\beta - \alpha), \tan \beta)$ parameter plane \cite{Ferreira2014,Bernon:2015wef,Basler:2017nzu,Ferreira:2017bnx}. {They can be understood by examining the behaviour of $\kappa_{hqq}$, the coupling between the SM-like Higgs $h$ and the quarks normalised to the corresponding SM value, as a function of the angles $\alpha$ and $\beta$. Taking the limits $\beta - \alpha\rightarrow \frac{\pi}{2}$ (upper lines) and $\beta + \alpha \rightarrow \frac{\pi}{2}$ (lower lines), the couplings become:}
{
\begin{equation}
\begin{aligned}
	\kappa_{hdd}
	=
	- \frac{\sin \alpha}{ \cos \beta}
	&=
	\sin (\beta  - \alpha) - \cos( \beta - \alpha) \tan \beta 
	\xrightarrow[\beta - \alpha = \frac{\pi}{2}]{}
	1 ~ \textrm{(middle-region),}
	\\ 
	&=
	- \sin (\beta + \alpha) + \cos(\beta + \alpha ) \tan \beta
	\xrightarrow[\beta + \alpha = \frac{\pi}{2}]{}
	- 1 ~ \textrm{(right-arm),}
	\\
	\kappa_{huu}
	=
	\frac{\cos \alpha}{ \sin \beta}
	&=
	\sin (\beta  - \alpha) + \cos( \beta - \alpha) \cot \beta
	\xrightarrow[\beta - \alpha = \frac{\pi}{2}]{}
	1 ~ \textrm{(middle-region),}
	\\
	&=
	\sin (\beta + \alpha) + \cos(\beta + \alpha ) \cot \beta 
	\xrightarrow[\beta + \alpha = \frac{\pi}{2}]{}
	1 ~ \textrm{(right-arm).}
\end{aligned}
\end{equation}
}

\begin{figure}[t]
\begin{center}
\includegraphics[width=0.40\linewidth]{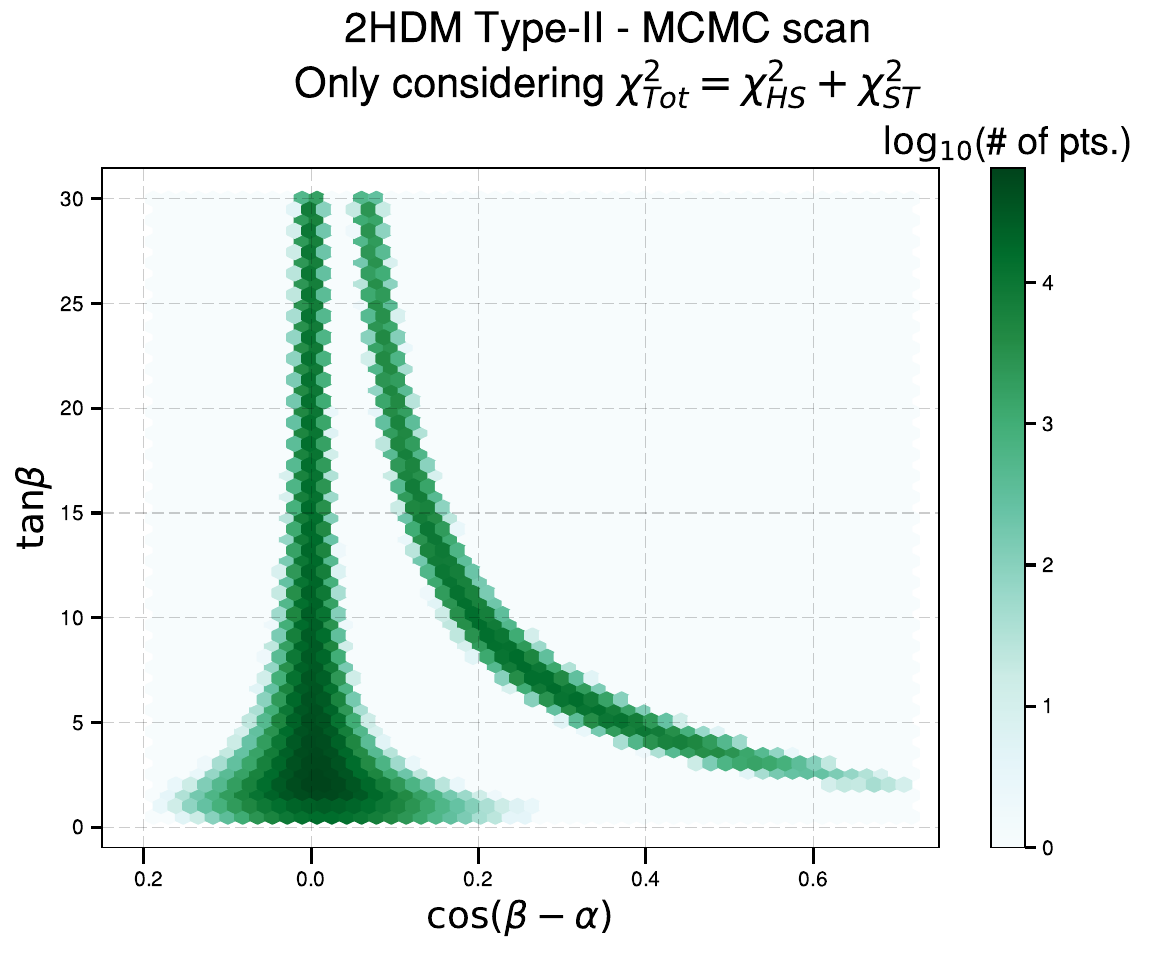}
\includegraphics[width=0.46\linewidth]{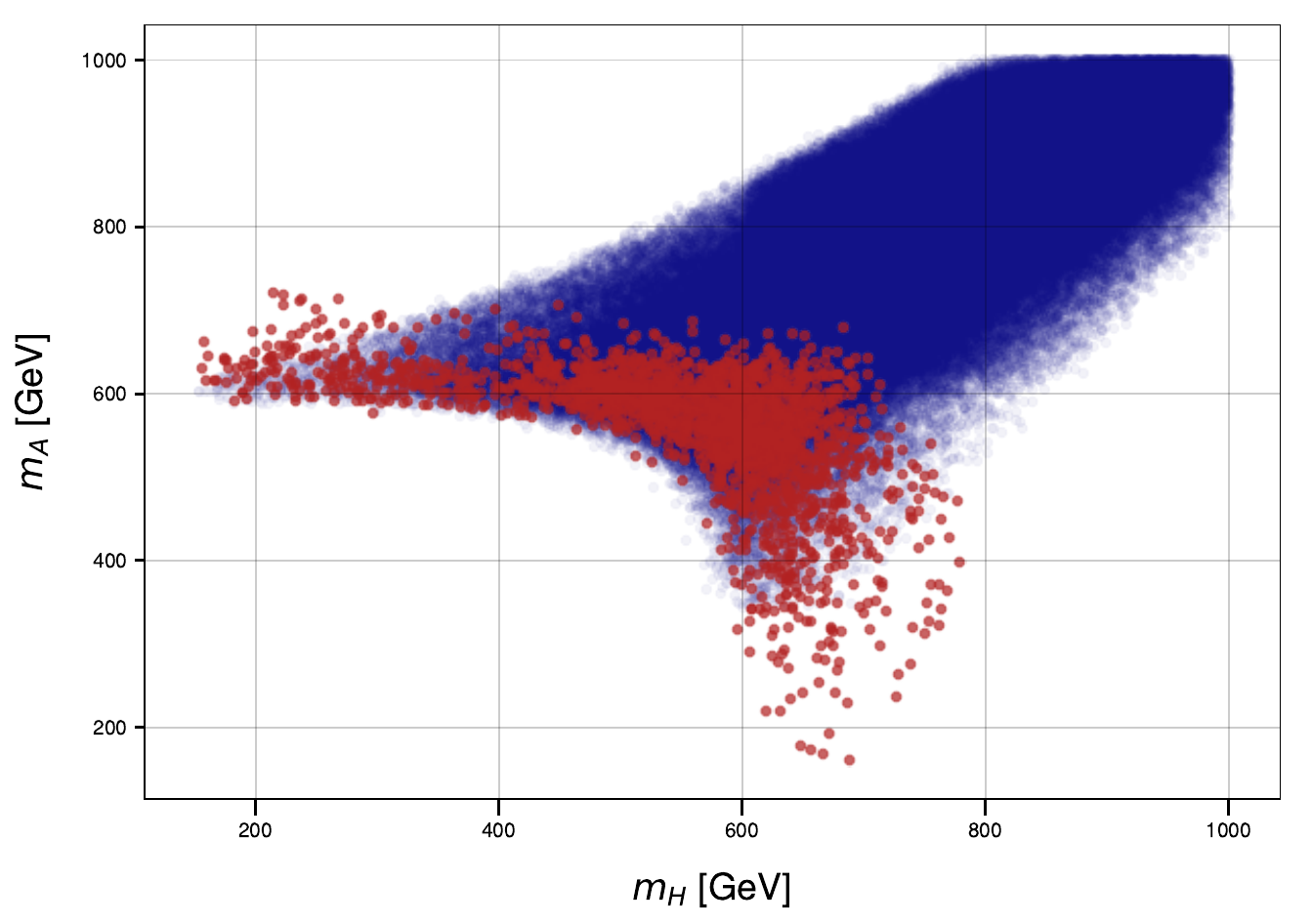}
\caption{Distribution of the 2HDM Type-II parameter space points, in the ($\cos (\beta - \alpha), \tan \beta$) plane (left) and in the $(m_H, m_A)$ plane (right), allowed by the current experimental and theoretical constraints  described in the text. The bound on the charged Higgs mass is implemented as $m_{H^\pm}\ge$ 600 GeV. In the right panel, the blue dots represent the alignment region while the red ones refer to the wrong-sign one.}
\label{fig:2HDm_HS_ST} 
\end{center}
\end{figure}

\noindent
{The ``middle-region'' (containing the aforementioned alignment limit of the 2HDM), which is the SM limit of the theory, is the region where the coupling $\kappa_{hdd}$ is positive. It is illustrated in Fig.~\ref{fig:2HDm_HS_ST} by the contour nearly symmetric around $\cos (\beta - \alpha ) = 0$. The  ``right-arm'', also called the wrong-sign scenario, is instead the region where the coupling $\kappa_{hdd}$ is negative. In the left-hand side plot of Fig.~\ref{fig:2HDm_HS_ST}, this region is represented by the narrow arm (or tongue) extending at large positive values of $\cos (\beta - \alpha )$.} 

\noindent
Recent studies from ATLAS \cite{ATLAS-CONF-2019-005} and CMS \cite{CMS-HIG-17-031-005} on the allowed regions of the 2HDM Type-II state that, although the data slightly prefer a positive sign of $\kappa_{huu}/\kappa_{hdd}$, the positive and negative hypotheses cannot be distinguished at the {95\%} Confidence Level (CL). On the theory side, an interesting study \cite{Basler:2017nzu} based on Renormalisation Group Equations (RGEs) has shown that, if one requires the model to be valid up to high energies (well beyond 1 \TeV), the allowed parameter space shrinks to the positive sign of $\kappa_{huu}/\kappa_{hdd}$. Below the \TeV energy scale, though, both the alignment limit and the wrong-sign scenario are possible. Hence, from a mere phenomenological point of view, it is not surprising to see that many analyses have been performed to constrain these two domains (see, e.g.,  Ref.~\cite{Ferreira:2017bnx} and references therein). 

\noindent
{Fig.~\ref{fig:2HDm_HS_ST} has been obtained considering bounds on the six free 2HDM Type-II parameters (recall that, here, $m_h=125$ GeV) coming from different sources. We refer to Ref.~\cite{Accomando:2019jrb} for the methodology employed to extract such constraints 
when taking into account the SM-like Higgs coupling strengths, null searches for new Higgs states, EWPOs and  theoretical constraints, all simultaneously.} In the left panel, we plot the allowed points in the plane ($\cos (\beta - \alpha), \tan \beta$), also displaying the density of these. In the right panel, we display the $(m_H, m_A)$ parameter space. The blue dots represent the alignment region while the red ones refer to the wrong-sign scenario. In both plots, we enforce the experimental bounds coming from HiggsSignals \cite{HiggsSignals} and HiggsBounds \cite{HiggsBounds1, HiggsBounds2, HiggsBounds3, HiggsBounds4}, EWPOs  plus the theoretical constraints from unitarity (upper bound at $8\pi$), perturbativity (upper bound at $4\pi$) and stability of the scalar potential. We moreover set the bound on the charged Higgs mass to be 600 GeV, as per constraints coming from $b\to s\gamma$ transitions \cite{Hpm}. In the right plot, one can see that, in the alignment limit of the 2HDM Type-II, the CP-odd Higgs state is required to be rather heavy: $m_A \ge 350$ GeV. Only in the wrong-sign scenario, it can in principle have a mass as light as $m_A \simeq$ 150 GeV (see red dots), when $Z_7$ is rather large and positive definite. The latter feature is the result of the effects coming from the  enforcement of the perturbativity and stability of the scalar potential. This picture depends however on the limit that could be in future set on the charged Higgs boson mass. Raising the $m_{H^\pm}$ limit pushes the lower bound on $m_A$ further up, in the alignment scenario. In the wrong-sign domain, though, one can still have light CP-odd Higgs masses at the price of stretching $Z_7$ towards large and positive values, $Z_7\ge 1$, typically \cite{Accomando:2019jrb}. (This is in agreement with the findings of Ref.~\cite{Bernon:2015wef}.)

\noindent
Having shown that in the wrong-sign scenario low $m_A$ masses are still allowed, we are now ready to discuss the possibility to detect such a light CP-odd Higgs boson at the LHC in the $Z^*h$ channel.

\section{Numerical results}

{
\begin{table}[t]
\begin{center}
\begin{tabular}{|c|c|c|c|c|c|}
\hline
$m_A$ (\GeV) & $m_{H^\pm}$ (\GeV) & $m_H$ (\GeV) & $\cos (\beta - \alpha )$ & $\tan\beta$ & ~$Z_7$ \quad \\
\hline
190 & 659 & 585 & 0.36 & 4.9 & 1.8 \\
\hline
200 & 628 & 594 & 0.28 & 6.4 & 1.3 \\
\hline
210 & 625 & 597 & 0.26 & 6.9 & 1.1 \\
\hline
\end{tabular}
\caption{The three benchmark points considered. From left to right, we list the CP-odd Higgs boson mass, the charged Higgs boson mass, the heavy CP-even Higgs boson mass as well as the values of $\cos (\beta - \alpha )$ and $\tan\beta$. For completeness, we include also the $Z_7$ value. The mass of the SM-like Higgs boson is fixed to be $m_h = 125$ \GeV.}
\label{tab:scenarios}
\end{center}
\end{table}
}

\begin{figure}[t]
\begin{center}
\includegraphics[width=0.25\textwidth]{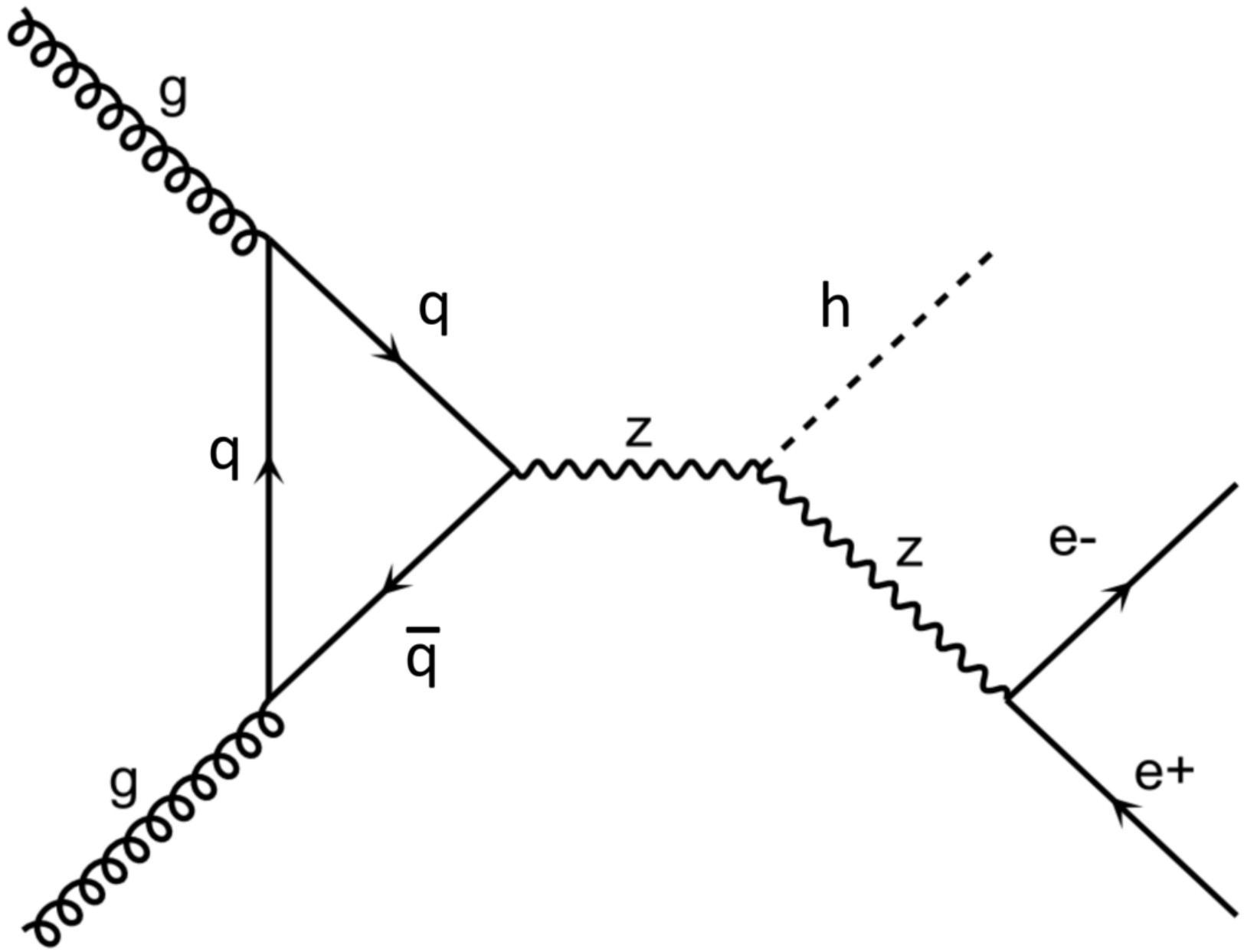}
\includegraphics[width=0.25\textwidth]{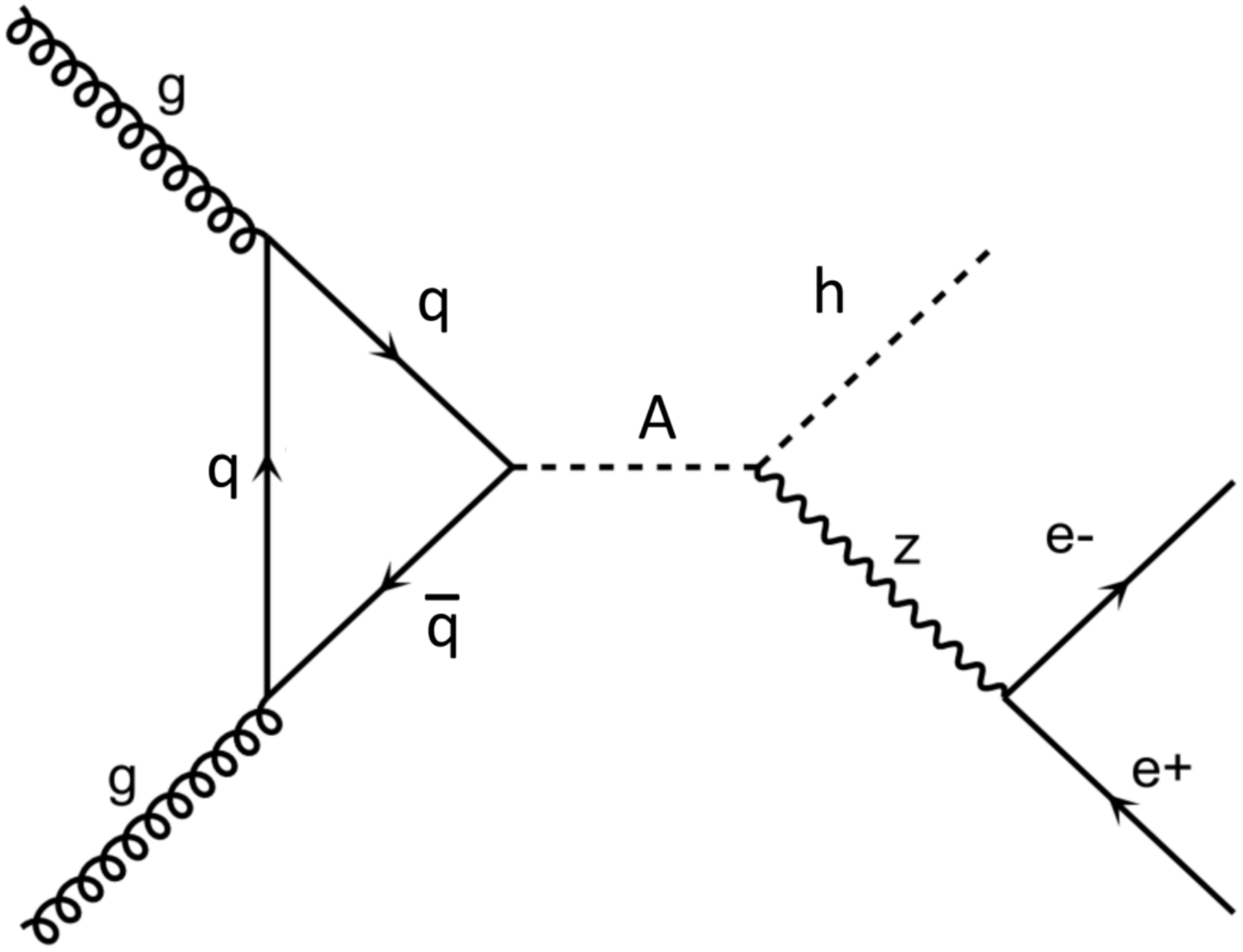}
\includegraphics[width=0.25\textwidth]{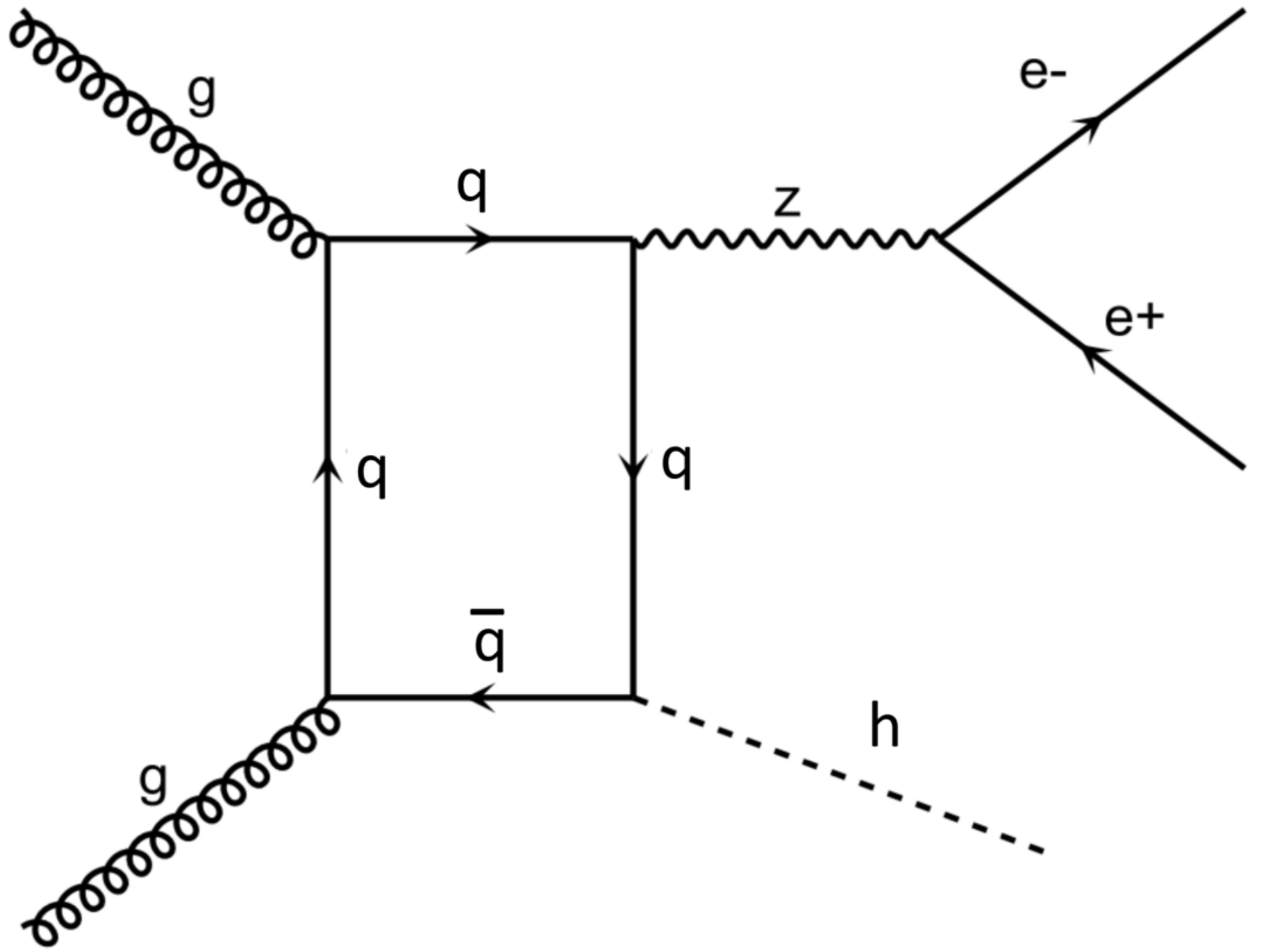}
\caption{Feynman diagrams for the process $pp\rightarrow Z^{(*)}h\rightarrow e^+e^-h$. The top row displays the quark-antiquark induced sub-process whereas the bottom row shows the gluon-gluon induced production sub-process.}
\label{fig:FD}
\end{center}
\end{figure}

\noindent
We consider the process $pp\rightarrow Z^*h \rightarrow l^+l^-h$ where the $Z$-boson decays into a lepton pair ($l = e, \mu )$. As the width of the SM-like Higgs boson is very small, we adopt the Narrow Width Approximation (NWA) and leave such a Higgs state on-shell. In this way, we have the freedom of multiplying our parton level results for the relevant Branching Ratios (BRs) of the Higgs boson. We include both the quark- and the gluon-induced sub-processes:
\begin{center}
$q\bar q\rightarrow Z^*h \rightarrow l^+l^- h$, \\
$        gg\rightarrow Z^*h \rightarrow l^+l^- h$. 
\end{center}
\noindent
 The Feynman graphs corresponding to the above sub-processes are visualised in Fig. \ref{fig:FD}.

\begin{figure}[t]
\begin{center}
\includegraphics[width=0.25\textwidth]{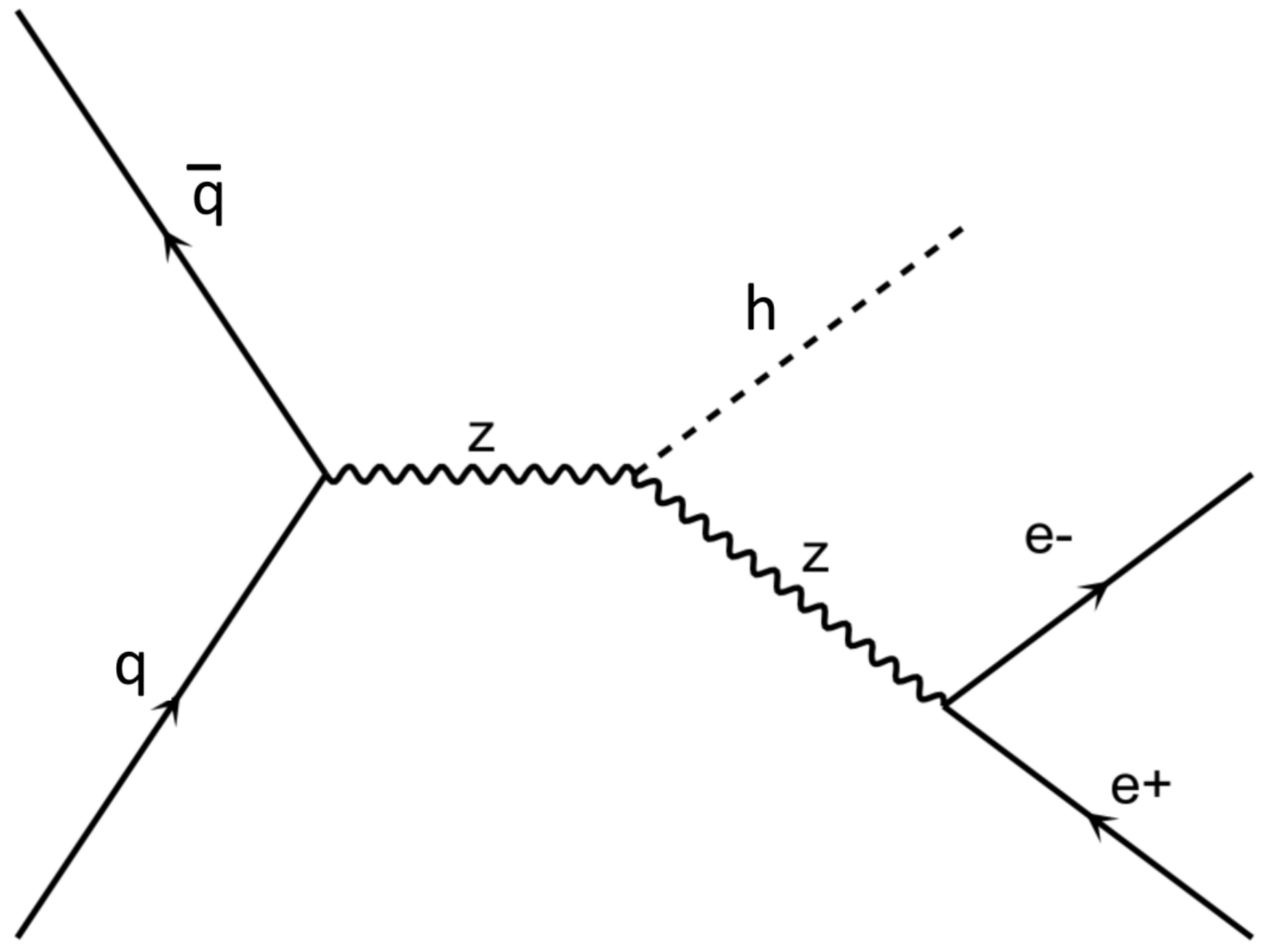}
\includegraphics[width=0.25\textwidth]{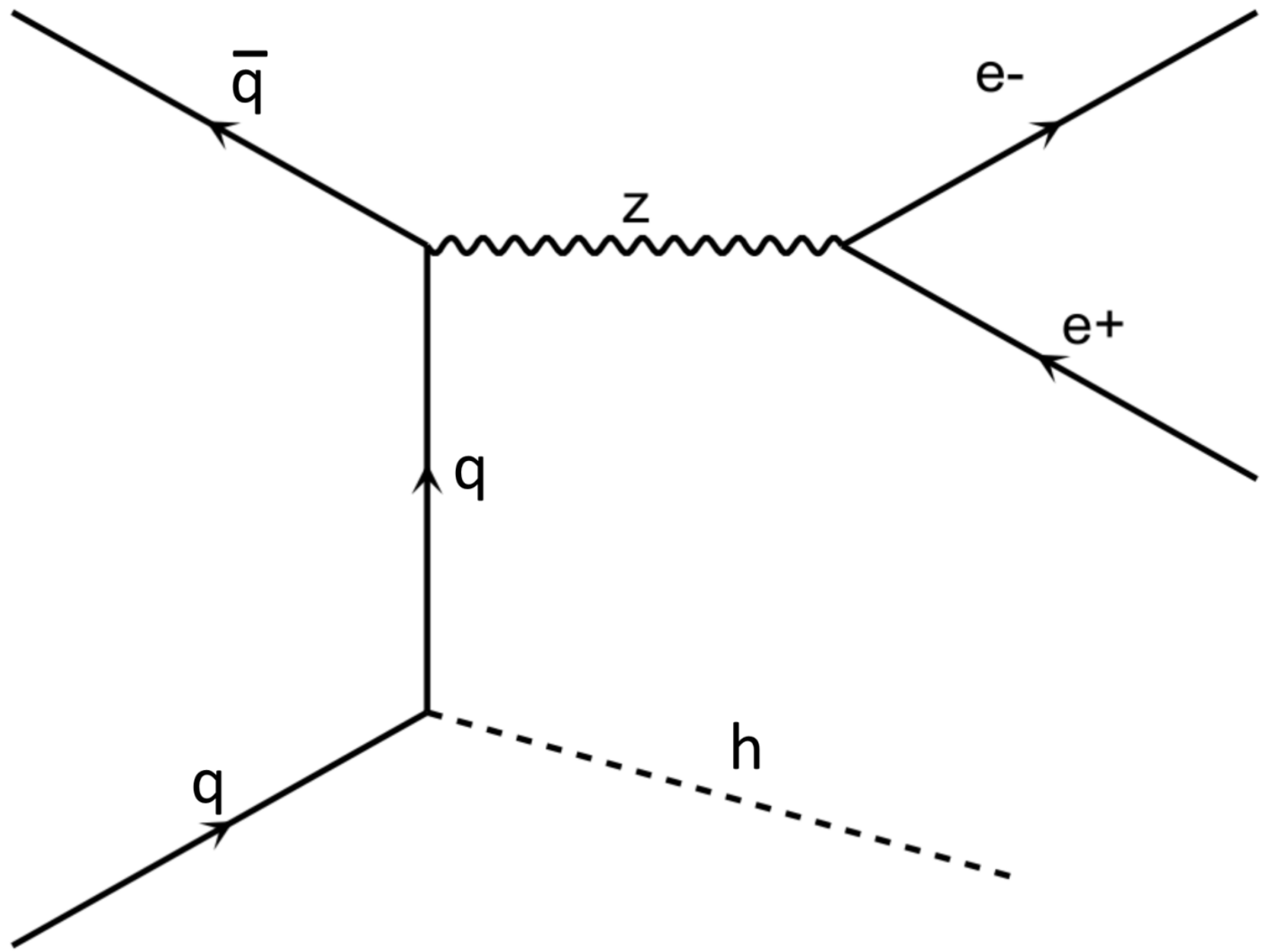}
\includegraphics[width=0.25\textwidth]{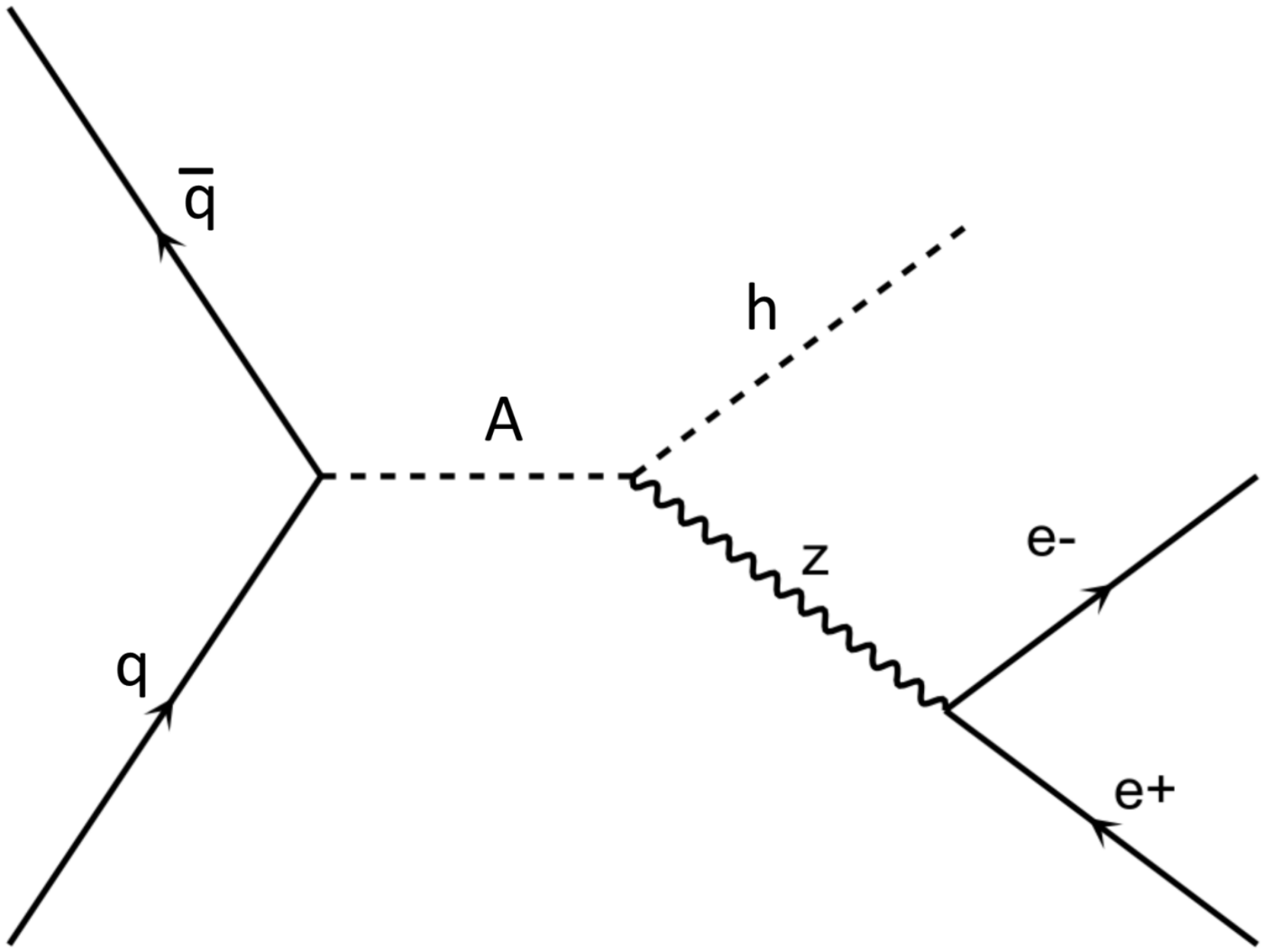}
\end{center}
\end{figure}

The aim of this paper is to show that a CP-odd Higgs boson with mass below the $Zh$ decay threshold, i.e.,   $m_A \le m_Z + m_h$, could still be observed at the  LHC. We therefore choose the input parameters given in Tab.\ref{tab:scenarios}. {These benchmark scenarios are representative of the low $A$ mass portion of the wrong-sign region.} If one indeed considers the aforementioned most recent ATLAS analysis of the process $pp\rightarrow A \rightarrow Zh\rightarrow Zb\bar b$ \cite{TheATLAScollaboration:2016loc}, one  notices that the CP-odd Higgs mass range starts at around $m_A = 220$ GeV. The search for the heavy CP-odd Higgs, $A$, decaying into a $Z$ boson and the 125 GeV Higgs state, is performed by looking at final states with either two opposite-sign charge leptons ($l^+l^-$ with $l = e, \mu$) or a neutrino pair ($\nu\bar\nu$) plus two $b$-jets at the 13 TeV LHC with a total integrated  luminosity $L = 3.2 $ fb$^{-1}$. 

\begin{figure}[t]
\centering 
\includegraphics[width=0.45\textwidth]{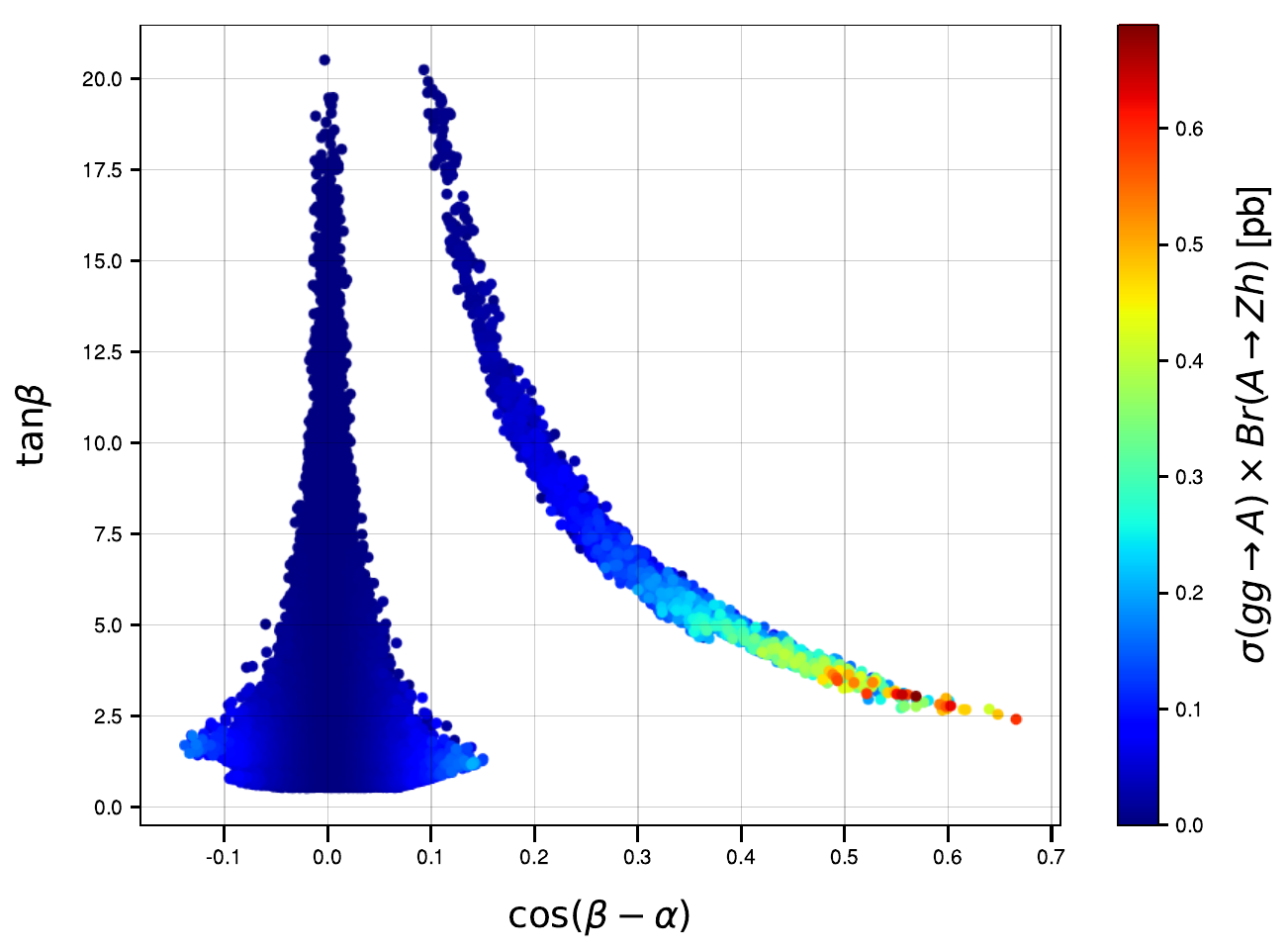}
\includegraphics[width=0.45\textwidth]{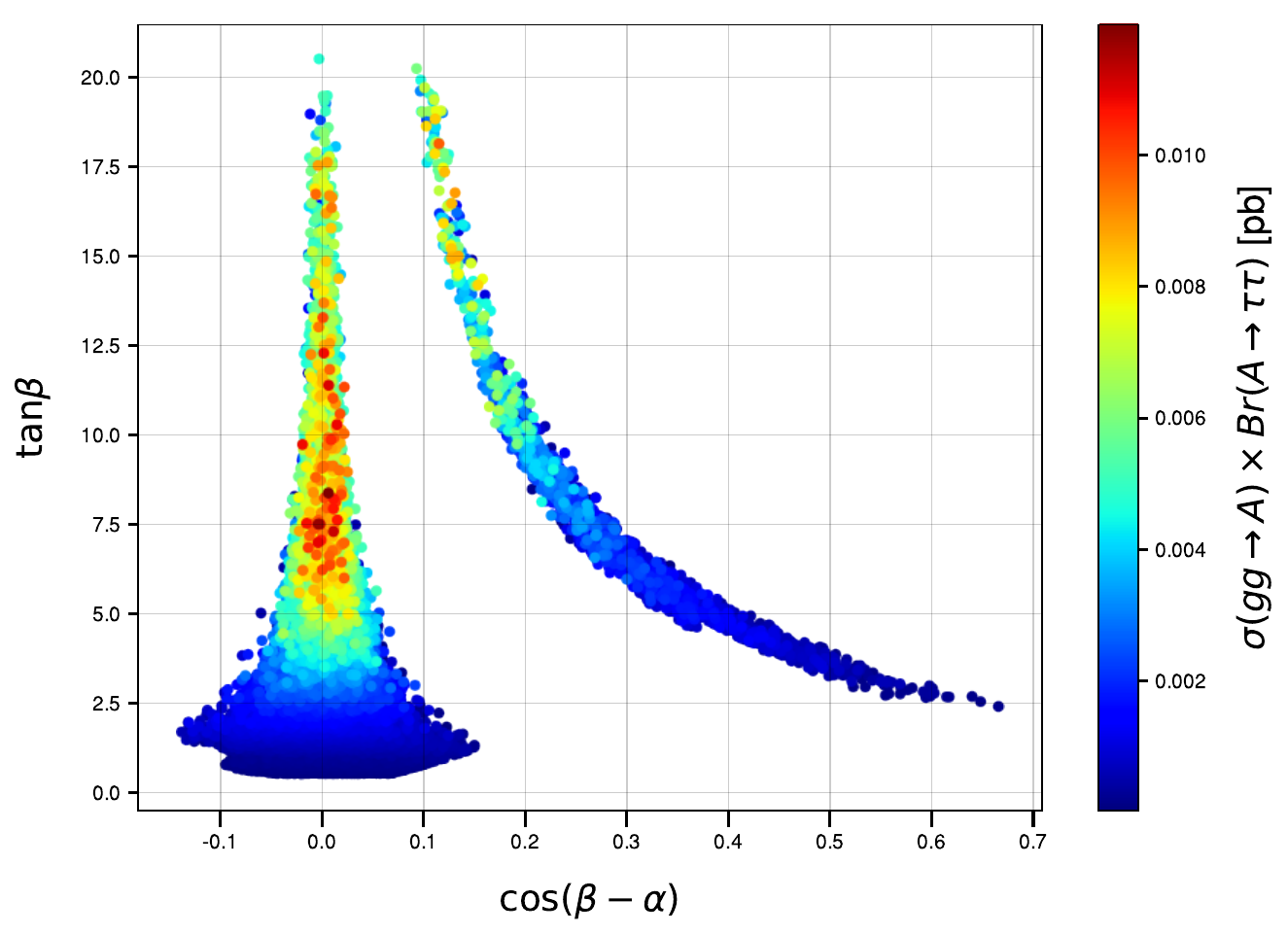}
\caption{Magnitude of the total cross-section times  BRs over the ($\cos (\alpha - \beta),
\tan\beta$) plane for two different processes mediated by the CP-odd Higgs state $A$. From left to right: $pp\rightarrow A\rightarrow Z^*h$ and $pp\rightarrow A\rightarrow\tau^-\tau^+$ (herein, we include only the gluon-gluon induced contribution).}
\label{fig:sigmaA}
\end{figure}

This analysis thus still leaves uncovered the below-threshold $A$  mass region. This part of the spectrum needs in fact to consider the $Z$ boson as being off-shell, in order to allow for the CP-odd Higgs boson to form a resonant peak. The present experimental analysis works instead under the approximation that the $Z$ boson is on-shell, i.e., it adopts the NWA for both the Higgs and  neutral gauge boson. The low mass region of the CP-odd Higgs state can be searched for in the $pp\rightarrow \tau^+\tau^-$ channel. However, this channel produces an enhanced cross section for medium-to-high values of $\tanb$. The $Z^*h$ channel that we are considering in this paper should be seen as complementary to that, as it gives raise to sizeable cross sections for low-to-medium  values of $\tanb$ and large $\cos (\alpha - \beta)$ where the $\tau$-channel is suppressed. This is shown in Fig.~\ref{fig:sigmaA}, where we plot the gluon-gluon induced cross-section times the decay BRs for the CP-odd Higgs boson over the ($\cos (\beta - \alpha)$,  $\tan\beta$) plane. The magnitude of the total rate is given following the colour code in the right columns.
This theoretical result finds confirmation in the experimental analysis performed by the CMS collaboration on their search for a CP-odd Higgs boson in the di-tau channel \cite{CMS-PAS-HIG-17-020}. There, they can go down to $\tan\beta\simeq 6$ to exclude masses in the range $m_A < 190$ \GeV at 95\% CL In this respect, our benchmark points are still viable and, if no new Higgs boson is found, could extend the exclusion region further down in 
$\tan\beta$ and up in $m_A$.  

\begin{figure}[t]
\begin{center}
\includegraphics[width=0.35\textwidth]{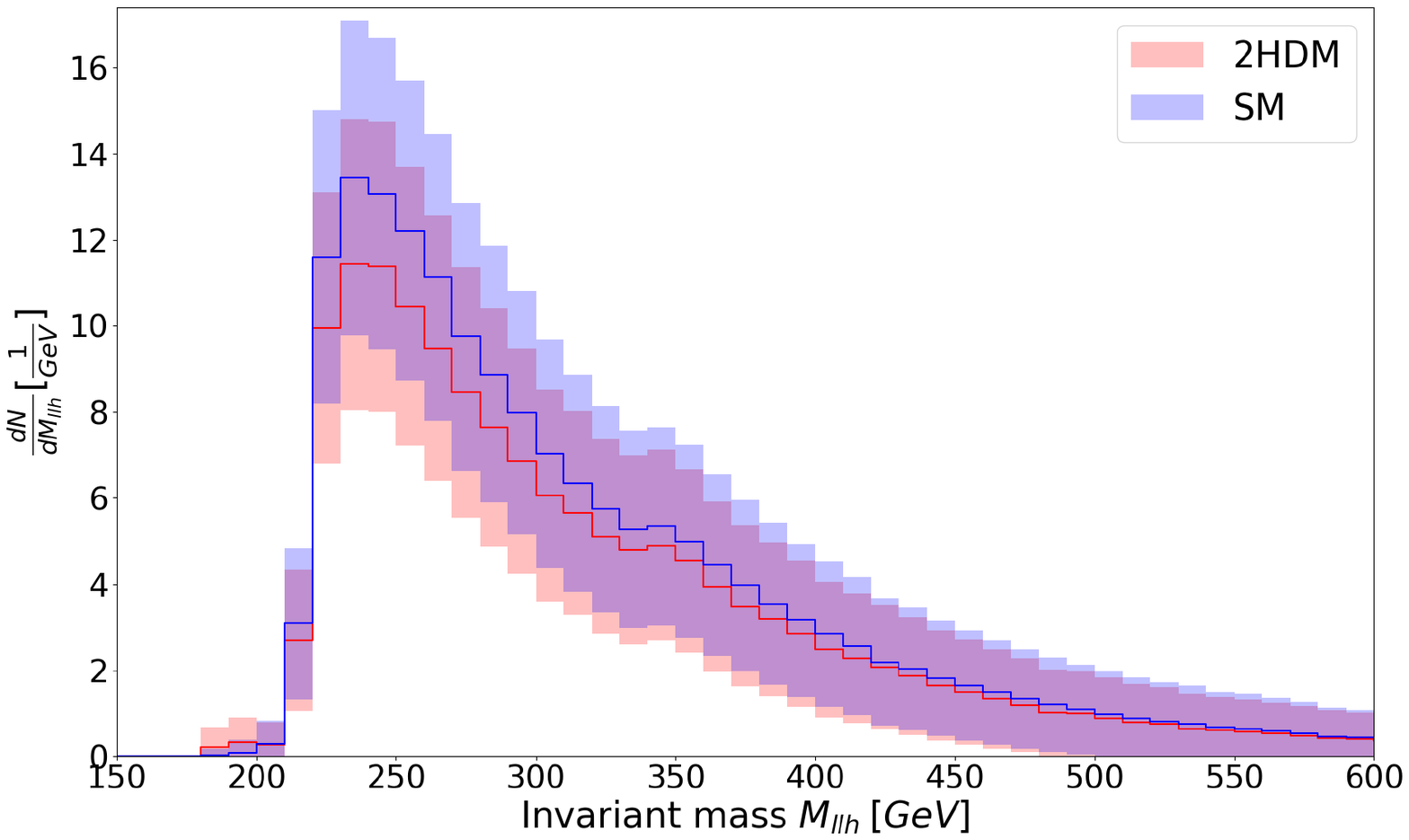}
\includegraphics[width=0.35\textwidth]{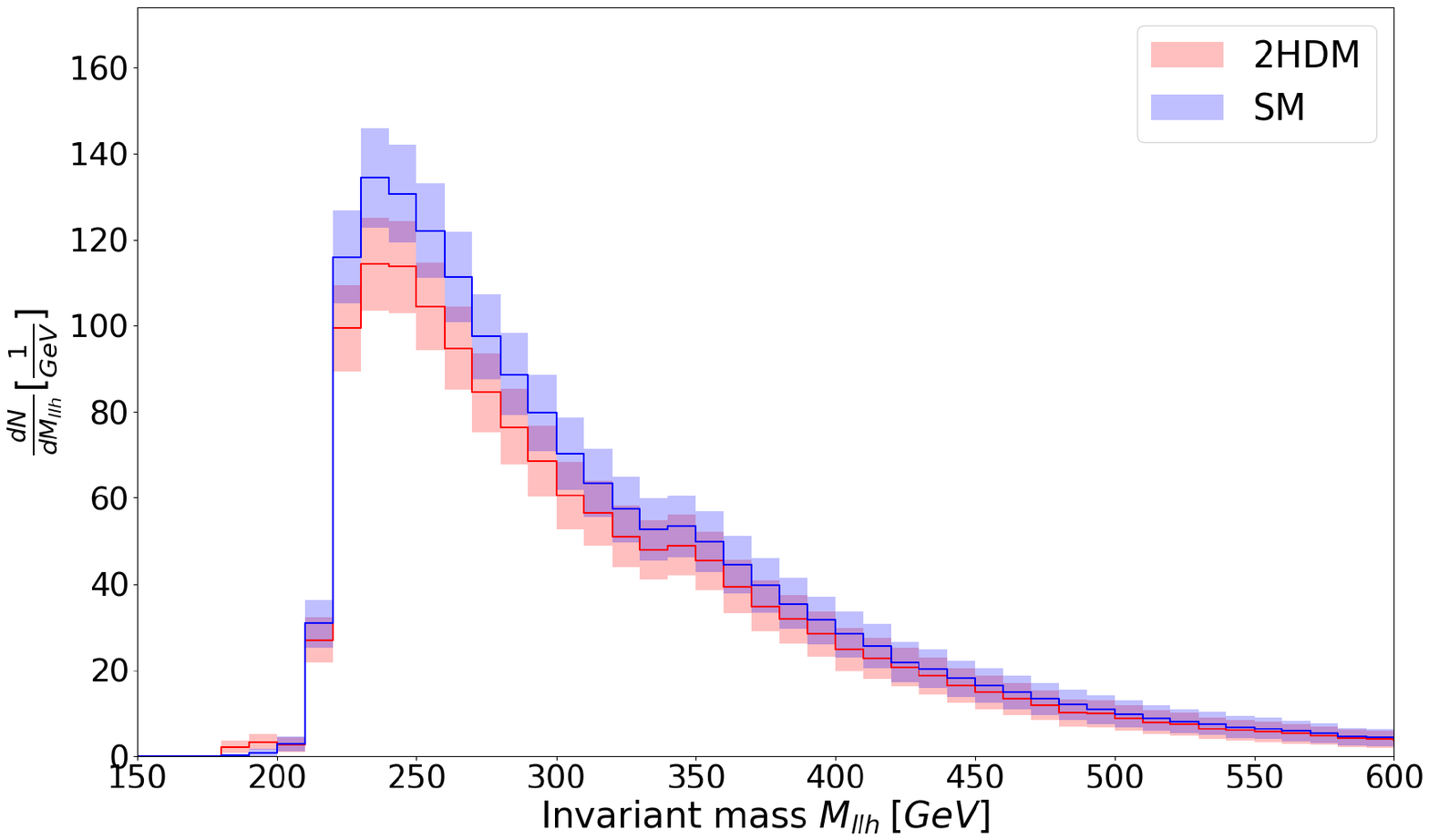}
\includegraphics[width=0.35\textwidth]{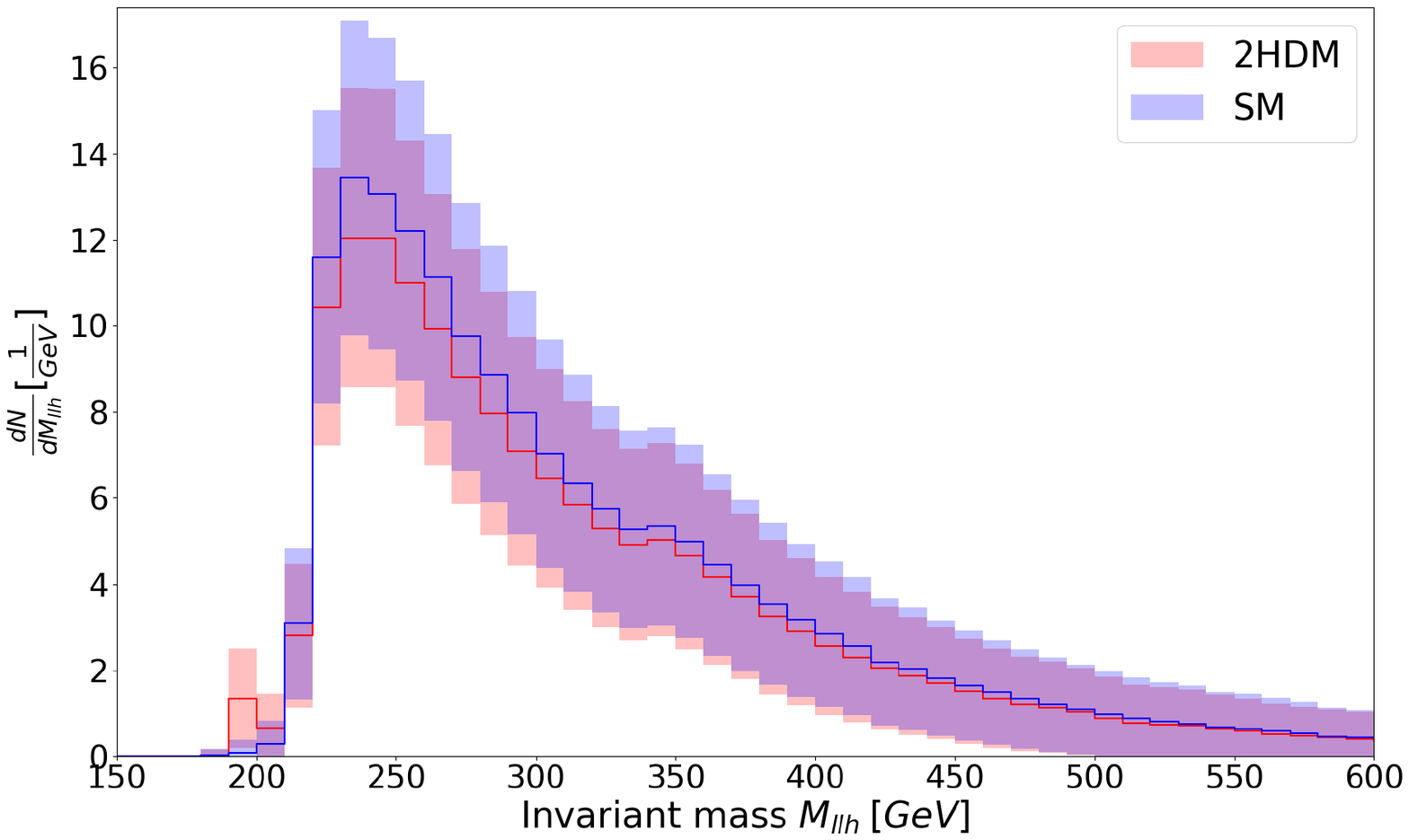}
\includegraphics[width=0.35\textwidth]{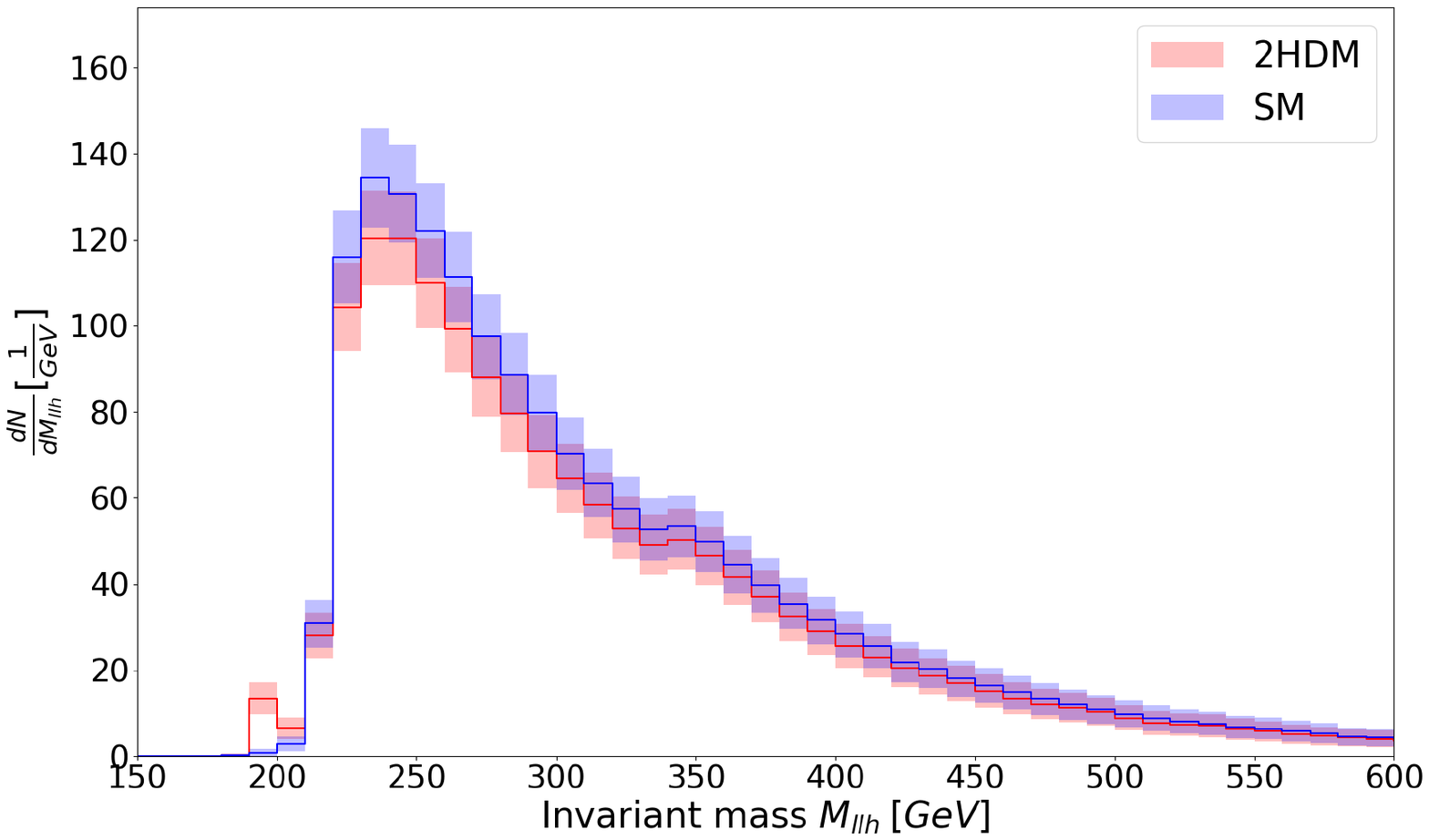}
\includegraphics[width=0.35\textwidth]{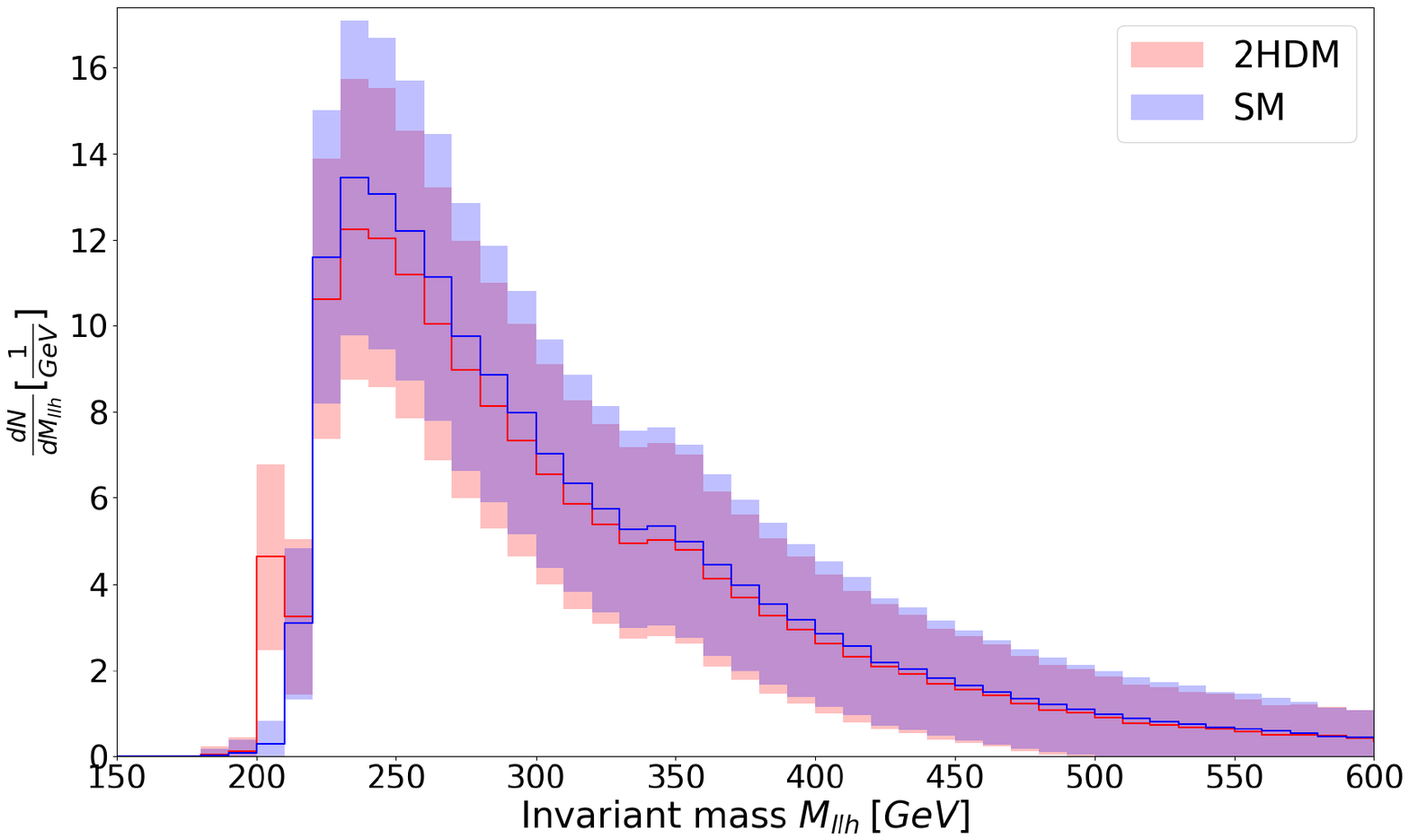}
\includegraphics[width=0.35\textwidth]{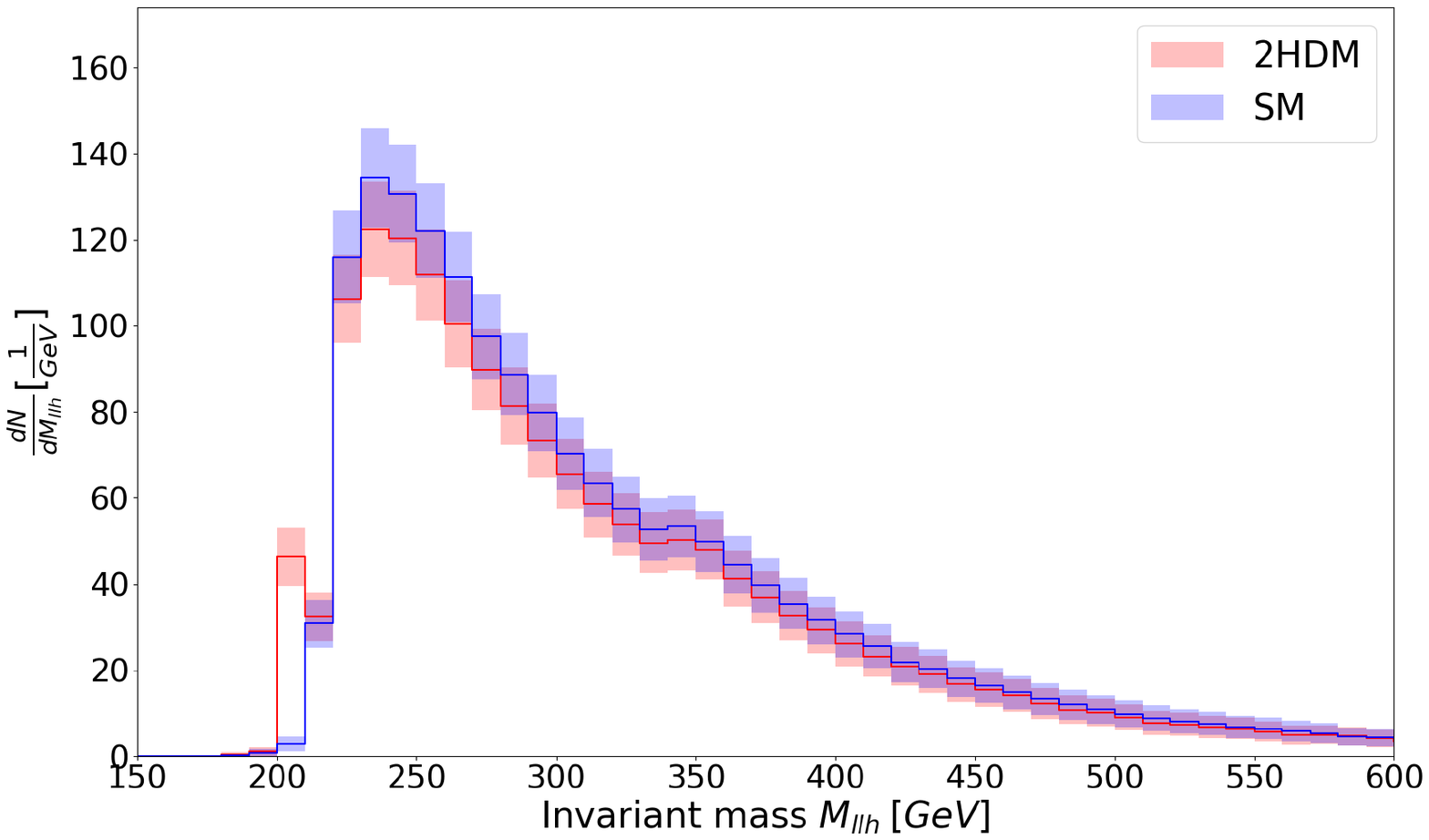}
\caption{Number of events in the invariant mass of the $l^+l^-h$ system for the three benchmark scenarios (red) as compared to the SM (blue). The binning is 10 \GeV. The shaded area is reflective of the statistical error. The top two graphs refer to the $m_A = 190$ GeV case, with the left-handed one evaluated at a luminosity $L = 100 ~{\rm fb}^{-1}$ and the right-handed one at $L = 1000 ~{\rm fb}^{-1}$. The middle graphs refer to the $m_A = 200$ GeV case and the bottom ones to the $m_A = 210$ GeV case (with the aforementioned luminosities from left to right).}
\label{fig:spectrum}
\end{center}
\end{figure}

\noindent
In our below-threshold analysis, we choose the specific channel $pp\rightarrow Z^*h\rightarrow l^+l^-b\bar b$, assuming the SM-like Higgs decay rate to be BR$(h\rightarrow b\bar b)=0.58$ in agreement with Ref. \cite{CMS-HIG-17-031-005}. We apply acceptance cuts on the charged leptons as in Ref.~\cite{TheATLAScollaboration:2016loc}: $|\eta_l|<2.5$ and $p^l_T>10$ \GeV. The NN23LO1 Parton Distribution Function (PDF) set is used in the five flavour scheme, corresponding to the strong coupling value of $\alpha_s(M_Z) = 0.13$ \cite{nn23lo1}. We have treated both signal sub-processes at Leading Order (LO) for consistency, as the $gg$ induced one cannot be computed at Next-to-LO (NLO), given that this would involve two-loop massive three- and four-point functions which are not available in the numerical framework adopted. However, the $b\bar b$ induced channel, being an ElectroWeak process at LO, can easily be computed at both NLO and Next-to-NLO (NNLO) in QCD. We have estimated the magnitude of such higher-order corrections using a variety of PDFs, including NNPDF2.3 NLO and NNPDF2.3 NNLO sets \cite{nn23lo1}. (Notice that we do not resolve the real QCD radiation from the initial state in our analysis.)
We find typical ${\cal O}(15\%)$ effects in agreement with Ref.~\cite{HAN1991167}. These corrections will therefore not alter our conclusions significantly.

\noindent
In order to quantify the theoretical uncertainties, which constitute an important fraction of the systematical error, we have computed the variation of the cross section with the renormalisation/factorisation scales for the two separate channels, $q\bar q$ and $gg$ induced. We have considered a set of three values for the above-mentioned scales: $Q_F = Q_R = \{0.5\times\sqrt{\hat s}, \sqrt{\hat s}, 2\times\sqrt{\hat s}\}$ where $\sqrt {\hat s} = M_{l^+l^-h}$. The outcome for the three benchmark scenarios in Tab.~\ref{tab:scenarios} is that the scale-dependence for the $q\bar q$ induced process is around $8\%$, while it increases up to order $60\%$ for the $gg$ induced process.The latter result is expected as the $gg$ process is much more sensitive to higher-order corrections than the $q\bar q$ one. However, the $gg$ contribution to the 2HDM signal is sub-dominant compared to the $q\bar q$ one. In the binned differential cross-section in the reconstructed $A$ mass, it’s below ${\cal O}(5\% )$ around the hypothetical CP-odd Higgs mass. Thus, the associated theoretical uncertainty does not affect the search. At higher invariant masses instead it becomes more sizeable.
In particular, at masses around the top pair threshold, $M_{l^+l^-h}\simeq$ 350\GeV, the $gg$ process generates a shoulder-shaped excess over the SM $q\bar q$ background. This contribution comes from the combined effect of the violation of the Furry theorem in the SM, due to the off-shellness of the intermediate $Z$-boson, the additional CP-odd Higgs contribution in the 2HDM and their interference. Here, the $gg$ contribution amounts to around $15\%$ of the total cross section. For high luminosities, when the LHC will gain sensitivity to such an excess, the computation of the higher-order corrections will become essential in order to have full control of the systematics. 

In the following, to display these results, the renormalisation and factorisation scales are set equal to $\sqrt {\hat s} = M_{l^+l^-h}$ (i.e., the center-of-mass energy at parton level) on on an event-by-event basis. For the benchmark points in Tab.~\ref{tab:scenarios}, which pass all experimental and theoretical constraints mentioned in the previous section, we compute the differential distribution in the reconstructed invariant mass of the $l^+l^-h$ system by using MadGraph5\_aMC@NLO \cite{Madgraph}. This distribution shows a resonant peak centred on $m_A$ and the overwhelming SM background given by the $Z$ boson mediated processes. The resonant peaking structure is dominated by the $b\bar b$ induced process, whereas the gluon fusion initiated sub-process gives its major contribution above the $t\bar t$ threshold, appearing as a broad shoulder standing over the SM background at around $M_{l^+l^-h} = 2 m_t = 350$ \GeV. The spectrum is shown in Fig.~\ref{fig:spectrum}, where it has been chosen a binning of 10 GeV corresponding to a mass resolution of the $l^+l^-h$ system of roughly 5\%. As one can see, the excess of events for $m_A = 190$ GeV (top two plots) stands nicely over a nearly null SM background, even if it cannot display a true peaking structure. Moreover, it is separated enough from the bulk of the SM background, which starts raising at around 210 GeV. As there is no clear resonant structure, the observation of such an excess of events heavily relies on the high-luminosity LHC run and would benefit from a restricted search window focused around that hypothetical value of the CP-odd Higgs mass. The right-hand side plot at $L = 1000~{\rm fb}^{-1}$ is in fact much more optimistic. The $m_A = 200$ GeV Breit-Wigner peak (two middle plots) is instead fully visible and, still, separated enough from the SM background on the right-hand side. This finding is quite promising for any potential search. With increasing  mass of the CP-odd Higgs boson, the signal gets even more enhanced and well shaped, a priori, as shown by the two bottom plots, where $m_A = 210$ GeV. However, considering parton shower, hadronisation and detector effects, which are beyond the scope of this letter, the risk for the signal to be more blurred increases as the $A$ peak would tend to flatten and overlap with the SM background. Despite this consideration, by tailoring the analysis so that the search window is centred around the potential mass of the CP-odd Higgs boson, the significance is encouraging in all the three cases. Tab.~\ref{tab:events} summarises the significance of the three benchmark points in different bins around the $A$ mass for a typical value of luminosity, e.g., $L = 100~{\rm fb}^{-1}$. In order to compare with the present experimental analyses, we have multiplied here the cross section in each bin by the mentioned BR of the SM-like Higgs into a $b\bar b$ pair.  The significance of a possible $A$  signal in the full final state $l^+l^-b\bar b$ remains quite good. Clearly, already analysing the present data from Run 2, one could have an evidence of a CP-odd Higgs boson with mass $m_A = 190$ GeV and a discovery of a slightly heavier $A$  with either $m_A = 200$ GeV or $m_A = 210$ GeV. By increasing the luminosity up to $L = 1000~{\rm fb}^{-1}$, a projected value at the High-Luminosity LHC (HL-LHC), the significance is expected to scale up by a factor of three, roughly. This would  then open up the possibility to claim a discovery of a below threshold $A$ decay in the discussed channel over the full mass range $190 ~\GeV\le m_A \le 210 ~\GeV$.

{
\begin{table}[t!]
\begin{center}
\begin{tabular}{|c|c|c|c|c|}
\hline
$bin$ & $N({\rm SM})$ & $N$($m_A = 190$ GeV) & $N$($m_A = 200$ GeV) & $N$($m_A = 210$ GeV) \\
\hline
180 & 0.2 & 2.1 ($\sigma = 2.6)$ & 0.2 ($\sigma = 0)$ & 0.4 ($\sigma = 0)$ \\
\hline
190 & 0.8 & 3.2 ($\sigma = 3.2)$ & 13.5 ($\sigma = 16.7)$ & 1.0 ($\sigma = 0)$ \\
\hline
200 & 2.8 & 2.7 ($\sigma = 0)$ & 6.5 ($\sigma = 2.3)$ & 46.2 ($\sigma = 25.7)$ \\
\hline
210 & 30.8 & 27 ($\sigma = 0.7)$ & 28 ($\sigma = 0.5)$ & 32.4 ($\sigma = 0.3)$ \\
\hline
\end{tabular}
\caption{Number of events for the full process $pp\rightarrow Z^*h\rightarrow l^+l^-b\bar b$ in four different bins for the three scenarios considered, at a luminosity $L = 100~ {\rm fb}^{-1}$. The significance in each bin is given by the number in brackets and it has been naively computed as $(N-N({\rm SM}))/\sqrt{N({\rm SM})}$ (if $N({\rm SM})\le 1$ then we divide by $N({\rm SM}) = 1$).}
\label{tab:events}
\end{center}
\end{table}
}

\section{Conclusions}

In summary, within the 2HDM Type-II, there exists at the LHC the possibility of accessing CP-odd Higgs boson signals  
via the processes $q\bar q, gg \to A\to Z^*h\to l^+l^- b\bar b$ ($l=e,\mu$), wherein the $Z$ boson is off-shell, thereby enabling sensitivity to $A$ masses below $m_Z+m_h\approx 215$ GeV, in fact, down to 200 GeV or so already at Run 2 and/or 3 while lower masses require HL-LHC data samples. This is an $m_A$ interval that is not being currently pursued by either ATLAS or CMS in this channel, so that we advocate the LHC experiments to investigate the signature we recommend. Indeed, a benefit of accessing this signal would also be the one of probing directly the so-called wrong-sign scenario of the 2HDM Type-II, an intriguing configuration, quite different from the alignment limit, as only in this case $m_A$ can be as light as the masses probed here.  

\section*{Acknowledgements}

\noindent
AM acknowledges the use of the IRIDIS High Performance Computing Facility and associated support services at the University of Southampton. EA and SM are supported in part through the NExT Institute and also acknowledge support from the STFC Consolidated grant ST/L000296/1. AM acknowledges support from the IDEX scholarship of the Université Paris-Saclay. All authors thank David Englert for considerable help at the beginning of this collaboration.


\end{document}